\documentclass[prx,floatfix,twocolumn,showpacs]{revtex4-1}
\usepackage{amsmath}
\usepackage{graphicx}
\usepackage{amsfonts}
\usepackage{amssymb}
\usepackage{bm}
\usepackage{epsfig,float,afterpage}
\usepackage[colorlinks,linkcolor=blue,citecolor=blue,urlcolor=blue]{hyperref}
\usepackage[usenames,dvipsnames]{color}
\usepackage{lipsum}
\def\beq{\begin{equation}}
\def\eeq{\end{equation}}
\def\bea{\begin{eqnarray}}
\def\eea{\end{eqnarray}}
\def\nn{\nonumber}
\def\ba{\begin{array}}
\def\ea{\end{array}}

\newcommand{\dg}{\dagger}

\newlength{\sizeonefig}
\newlength{\sizetwofig}
\begin{document}
\title{Dynamics of hybrid junctions of Majorana wires}
\author{Nilanjan Bondyopadhaya$^1$ and Dibyendu Roy$^2$}
\affiliation{$^1$ Integrated Science Education and Research Centre,  Visva-Bharati University, Santiniketan, WB 731235,  India}
\affiliation{$^2$ Raman Research Institute, Bangalore 560080, India}

\begin{abstract}
We investigate the dynamics of hybrid junctions made of the topological superconductor (TS) and normal metal (N) wires. We consider an X-Y-Z configuration for the junctions where X, Y, Z = TS, N. We assume the wires X and Z being semi-infinite and in thermal equilibrium. We connect the wires X and Z through the short Y wire at some time, and numerically study time-evolution of the full device. For TS-N-TS device, we find a persistent, oscillating electrical current at both junctions even when there is no thermal or voltage bias, and the TS wires are identical. The amplitude and period of the oscillating current depend on the initial conditions of the middle N wire indicating the absence of thermalization. This zero-bias current vanishes at a long time for any of X and Z being an N wire or a TS wire near a topological phase transition. Using properties of different bound states within the superconducting gap, we develop a clear understanding of the oscillating currents.

\end{abstract}
\vspace{0.0cm}
\maketitle

Recent progress in search of Majorana zero modes (MZMs) in condensed matter systems has generated massive interest for better understanding, control, and engineering of such systems featuring these exotic quasiparticles \cite{AliceaReview2012,Beenakker2013,Stanescu2013,Elliott2015}. The MZMs have been theoretically proposed to emerge at the edges of a topological superconductor (TS) wire and to exhibit non-Abelian particle statistics \cite{Kitaev2001,LutchynPRL2010,OregPRL2010}. Several experiments have observed significant evidence of MZMs in electrical transport measurements with engineered TS wires \cite{MourikScience2012,DasNature2012,NadjPergeScience2014,AlbrechtNature2016}. These TS wires hosting Majorana quasiparticles are expected to be an essential component of future quantum devices such as fault-tolerant quantum computers. Therefore, it is necessary to investigate various properties including thermalization, transport and braiding operations in hybrid junctions of TS and normal metal (N) wires.

In the past ten years, there are many theoretical studies for electrical transport in different junctions of TS and N wires \cite{Beenakker2013,Stanescu2013}. Most of these studies discuss steady-state transport to derive zero-bias conductance \cite{BolechPRL2007,LawPRL2009,Flensberg2010, RoyPRB2012, RoyPRB2013}, current-voltage characteristics \cite{RoyPRB2012,Zazunov2016}, equilibrium current-phase relation for fractional Josephson effect \cite{Rokhinson2012,Zazunov2012,Zazunov2016,Ioselevich2016} and robust Majorana conductance peaks \cite{RubyPRL2015,Yang2015,Sharma2016}. There are also some time-evolution study after a quantum quench \cite{Vasseur2014, Hegde2015, Sacramento2016} such as when an N wire is suddenly connected to a TS wire. Interestingly, a systematic study for the dynamics of tunneling current in these hybrid devices is missing. In this Letter, we study transient and steady-state quantum transport in hybrid junctions of X-Y-Z configuration where X, Y, Z = TS, N. We primarily address route to equilibration in these junctions by investigating the dynamics of tunneling current. Ideally, we consider the wires X and Z to be semi-infinite and in thermal equilibrium. However, we can only investigate $L_{X,Z}/L_Y >>1$  in our numerics for time-evolution where $L_{X,Z}$ and $L_Y$ are the length of the wire X or Z and Y respectively. We connect the wires X and Z through the middle Y wire at some time $t_0$ and explore time-evolution of the full system for various initialization of the middle wire. We probe electrical current through the X-Y and Y-Z junctions at every stage of time-evolution. Here we do not consider a voltage bias which can be incorporated through time-dependent phases in the tunnel couplings across the junctions \cite{Zazunov2016}. We also ignore any effect which can arise due to phase difference across the junctions of TS wires; therefore we set the phase of the TS wires to be same (or zero for simplicity) everywhere. 

We find a non-decaying, oscillating electrical current at both junctions of a TS-N-TS device even in the absence of any thermal or voltage bias and for identical TS wires (no phase difference). The amplitude and period of the oscillating current depend on the energy eigenstates of the wires which are localized around the junctions and within the superconducting gap of the TS wires, and on the initial conditions of the N wire. Since the wave functions of the energy eigenstates of the N wire in the superconducting gap decay exponentially deep inside the TS wire, they act as bound states. The presence of such bound states prevent equilibration (thermalization) of the middle N wire with the boundary wires \cite{DharPRB2006}, and this results into the non-uniqueness (initial-condition dependence) of the persistent current. There is no zero-bias  oscillating current at long time in an N-TS-N and an X-N-Z device when any of the X and Z being an N wire or a TS wire near a topological phase transition. In these latter cases, the energy spectrum of one of the boundary wire is gapless. Thus, there is no more bound state from the middle N wire, and the middle wire gets equilibrated with the boundary wire(s). Using properties of different bound states within the superconducting gap in our devices, we develop a theory to reproduce the simulated oscillating current with little numerics.

We now introduce the Hamiltonian for various components of the hybrid junctions. We model the N wire as a noninteracting tight-binding chain of spinless fermions and the TS wire by the Kitaev chain of a spinless p-wave superconductor \cite{Kitaev2001}. We write below a general Hamiltonian $H_{\rm \alpha}$ which can be used to represent N and TS wires by tuning the parameters:  
\bea
H_{\rm \alpha}&=&-\gamma_{\rm \alpha}\sum_{l'=1}^{L_{\rm \alpha}-1}(c^{\dg}_{l'}c_{l'+1}+c^{\dg}_{l'+1}c_{l'})+\epsilon_{\rm \alpha} \sum_{l'=1}^{L_{\rm \alpha}}c^{\dg}_{l'}c_{l'}\nn\\&+&\Delta_{\rm \alpha}\sum_{l'=1}^{L_{\rm \alpha}-1}(c^{\dg}_{l'}c^{\dg}_{l'+1}+c_{l'+1}c_{l'}),\label{Ham}
\eea
where $c_{l'}~(c^{\dg}_{l'})$ indicates annihilation (creation) operator of a spinless fermion at site ${l'}$ of the wire segment $\alpha=X,Y,Z$. Here, the parameters $\gamma_{\rm \alpha},\epsilon_{\rm \alpha}$ and $\Delta_{\rm \alpha}$ denote respectively hopping, on-site energy and superconducting pairing energy, and we take them to be real. The Hamiltonian $H_{\rm \alpha}$ indicates a N wire in the absence of pairing ($\Delta_{\rm \alpha}=0$). In the presence of pairing $\Delta_{\rm \alpha} \ne 0$, the superconducting wire undergoes a topological phase transition as $\epsilon_{\rm \alpha}$ is increased across $2\gamma_{\rm \alpha}$. The wire is in a topological phase for $\epsilon_{\rm \alpha}<2\gamma_{\rm \alpha}$ and the TS wire hosts two MZMs at the opposite ends of the wire for a relatively long wire. The wire is in a topologically trivial phase without the MZMs for $\epsilon_{\rm \alpha}>2\gamma_{\rm \alpha}$. The topological phase transition near $\epsilon_{\rm \alpha}=2\gamma_{\rm \alpha}$ is also accompanied by a bulk-gap closing. The superconducting wire has a large bulk-gap in its spectrum both in the topologically non-trivial and trivial phases, and the gap vanishes at the topological phase transition around $\epsilon_{\rm \alpha}=2\gamma_{\rm \alpha}$.

The tunneling Hamiltonian for the X-Y and Y-Z junction is independent of nature of X, Y, Z, and we take it of the following form:
\bea
H_{\alpha\beta}=-\gamma_{\rm{\alpha \beta}}(c_{l'}^{\dg}c_{l'+1}+c_{l'+1}^{\dg}c_{l'}),
\eea
where $\alpha\beta={\rm XY}$, $l'=L_X$, and $\alpha\beta={\rm YZ}$, $l'=L_X+L_Y$ respectively for the X-Y and Y-Z junction. For simplicity, we assume here the tunneling rate $\gamma_{\alpha \beta}$ (with $\gamma_{\alpha \beta} \ll \gamma_{\alpha}, \gamma_{\beta}$) to be same for both the junctions, i.e., $\gamma_{XY}=\gamma_{YZ}=\gamma'$. The full Hamiltonian of the hybrid device is $H_{\rm F}=H_{\rm X}+H_{\rm Y}+H_{\rm Z}+H_{\rm XY}+H_{\rm YZ}$. The full device consists of $L=L_X+L_Y+L_Z$ number of fermionic sites. Using conservation of electrical charges across the junctions, we define the electrical current at the junctions as
\bea
J_{\rm \alpha \beta}=-i\gamma_{\rm{\alpha \beta}}\langle (c_{l'}^{\dg}c_{l'+1}-c_{l'+1}^{\dg}c_{l'})\rangle,\label{current}
\eea
where again $\alpha\beta={\rm XY}$, $l'=L_X$, and $\alpha\beta={\rm YZ}$, $l'=L_X+L_Y$ respectively for the X-Y and Y-Z junction. The expectation $\langle .. \rangle$ defines averaging over the initial states of the wires.

Due to the pairing term $\Delta_{\alpha}$ in the superconducting parts of the hybrid device, it is convenient to use a basis ${\bf a}\equiv(a_1, a_2, \dots, a_{2L-1}, a_{2L})^T=(c_1, c^{\dg}_1, \dots, c_{L}, c^{\dg}_{L})^T$ to write the quadratic Hamiltonian $H_{\rm F}={\bf a}^{\dg}\mathcal{H}^{\rm F}{\bf a}=\sum_{l,m}\mathcal{H}_{lm}^{\rm F}a^{\dg}_la_m$ in terms of the matrix $\mathcal{H}_{lm}^{\rm F}$. Thus, $a_{2l}=a^{\dg}_{2l-1}$. Clearly the index $l$ in $a_l$ (or $a_l^\dagger$) does not represent actual physical site of the wire. One can define a map to the physical site $l'$ of spinless fermions as: $l'=(l+1)/2$ for odd values of $l$, and $l'=(l/2)$ for even values of $l$. We consider the wires X and Z are in thermal equilibrium at temperatures $T_{\rm X,Z}$ and chemical potentials $\mu_{\rm X,Z}$ before we connect them through the middle wire Y at time $t_0$. For isolated X and Z  wires, we have $H_{\alpha}=\sum_{l,m}\mathcal{H}_{lm}^{\alpha}a^{\dg}_la_m$ with $l,m=1,2,\dots,2L_{\rm X}$ for $\alpha={\rm X}$, and $l,m=2(L_X+L_Y)+1,2(L_X+L_Y)+2,\dots,2L$ for $\alpha={\rm Z}$. Therefore, we can write
\bea
\sum_{m}\mathcal{H}_{lm}^{\alpha}\psi^{\alpha}_q(m)=\lambda^{\alpha}_q\psi^{\alpha}_q(l),
\label{bathham}
\eea
where $\psi^{\alpha}_q(m)$ and $\lambda^{\alpha}_q$ represent the eigenvectors and eigenvalues of the wire $\alpha=X,Z$. Thus, the equilibrium density matrix for sites on the isolated boundary wires 
\bea
\langle a^{\dagger}_{l}(t_0) a_{m}(t_0) \rangle = \sum_q {\psi_q^{\alpha}}^*(l)\psi_q^{\alpha}(m) f(\lambda^{\alpha}_q,\mu_{\alpha},T_{\alpha}),\label{indm1}
\eea
with $l,m=1,2,\dots,2L_{\rm X}$ for $\alpha={\rm X}$ and $l,m=2(L_X+L_Y)+1,2(L_X+L_Y)+2,\dots,2L$ for $\alpha={\rm Z}$. Here, $f(\lambda^{\alpha}_q,\mu_{\alpha},T_{\alpha})=1/(e^{(\lambda^{\alpha}_q- \mu_{\alpha})/k_BT_{\alpha}}+1)$ is the Fermi function.

We also assume that the operators from different X, Y, Z wires are uncorrelated when they are disconnected at $t_0$. Therefore, we have 
\bea
&&\langle a^{\dagger}_{l}(t_0) a_{m}(t_0) \rangle =\langle a_{l}(t_0) a^{\dagger}_{m}(t_0)\rangle\nn\\&&=\langle a_{l}(t_0) a_{m}(t_0) \rangle=\langle a_{l}^{\dagger}(t_0) a^{\dagger}_{m}(t_0)\rangle=0 \,,\label{indm2}
\eea
where $l$ and $m$ represent indices corresponding to two different wires of the hybrid device. We here do not take a thermal distribution for the middle wire Y at $t_0$. Instead, we choose some arbitrary initial density matrix of wire Y, such as, 
\bea \langle a_{2l'}(t_0) a_{2m'-1}(t_0) \rangle = \left \{
\begin{array}{l l} n_{l'} & \quad \mbox{when}~l'=m' \\ 0 & \quad \mbox{when}~ l' \ne m' \end{array} \right. \label{indm3}
\eea
for physical sites: $l',m' \in \{L_{\rm X}+1,\dots, L_{\rm X}+L_{\rm Y}\}$. We wish to check whether the density matrix of the full device in the long-time limit $t\to\infty$ becomes independent of the initial density matrix of the middle wire. Such independence would indicate thermalization of the Y wire by the X and Z wires. We carry out this job by calculating time-evolution of electrical current at the junctions. So, we connect the wires by the tunneling Hamiltonians at time $t_0$ and study the time-evolution of the full device using the Heisenberg equations of motion. The solution of these equations of motion is given by
\bea
{\bf a}(t)=i \mathcal{G}^+(t-t_0){\bf a}(t_0),
\eea
where $\mathcal{G}^+(\tau)=-i e^{-i H_{\rm F} \tau} \theta (\tau)$ is the Green's function of the full device. Here, $\theta (\tau)$ is the Heaviside step function. Suppose, $\Psi_q(m)$ and $\Lambda_q$ denote the eigenvectors and eigenvalues of the full Hamiltonian $H_{\rm F}$. Therefore, 
\beq
\sum_{m=1}^{2L} \mathcal{H}^{\rm F}_{lm}\Psi_q(m)=\Lambda_q \Psi_q(l),~~l=1,2,\dots,2L.
\eeq
We apply the above relations to write the full Green's function in the following form for $t >t_0$:
\beq
\mathcal{G}^+_{r,s}(t-t_0)=- i \sum_{q=1}^{2L} \Psi_{q}(r) \Psi_{q}^*(s) e^{- i \Lambda_q (t-t_0)} \, ,
\label{EGF}
\eeq
where $r,s \in \{1,\dots, 2L\}$. Now, we can write the time-evolved density matrix of the full device as
\bea
\langle a^\dagger_l(t) a_m(t) \rangle= \sum_{r,s=1}^{2L} \mathcal{G}^+_{m,s}(t-t_0) \langle a^\dagger_{r}(t_0) a_{s}(t_0) \rangle [\mathcal{G}^+_{l,r}(t-t_0)]^{\dagger},\nn\\ \label{tdm}
\eea
where we plug the initial density matrix $\langle a^\dagger_{r}(t_0) a_{s}(t_0) \rangle$ from Eqs.~\ref{indm1},\ref{indm2},\ref{indm3}. Using Eq.~\ref{tdm} in Eq.~\ref{current}, we evaluate the time evolution of the electrical current at the junctions.

In the following numerical analysis, we fix $\gamma_{\rm X}=\gamma_{\rm Z}=\gamma >\gamma_{\rm Y}$; so that the bands of the boundary wires are broader than that of the middle. An N-N-N device in the above set-up has been studied in details in Ref.~\cite{DharPRB2006}, and it has been shown that there is a unique nonequilibrium steady-state (independent of initial values of $n_{l'}$) in the device when there is no single particle bound state from the middle wire (the band of wire Y lies within that of X and/or Z).

First, we investigate dynamics of an N-TS-N device whose steady-state transport characteristics are extensively studied both theoretically \cite{BolechPRL2007,LawPRL2009,Flensberg2010,RoyPRB2012, RoyPRB2013} as well as experimentally \cite{MourikScience2012,DasNature2012, AlbrechtNature2016} for detection of MZMs \footnote{An N-TS junction is only considered in many of these studies.}. We here prepare the decoupled N wires of our N-TS-N device initially in thermal equilibrium. We find from our numerics \cite{Suppl} that the nonequilibrium steady-state transport in the  N-TS-N device seems to be independent of the initial conditions of the finite TS wire when the band of the N wires is wider than that of the TS wire. Our present finding of unique nonequilibrium steady-state in the N-TS-N device validates all those previous steady-state transport analyses \cite{RoyPRB2012, RoyPRB2013}  in this system.             

\begin{figure}
\includegraphics[width=0.99\linewidth]{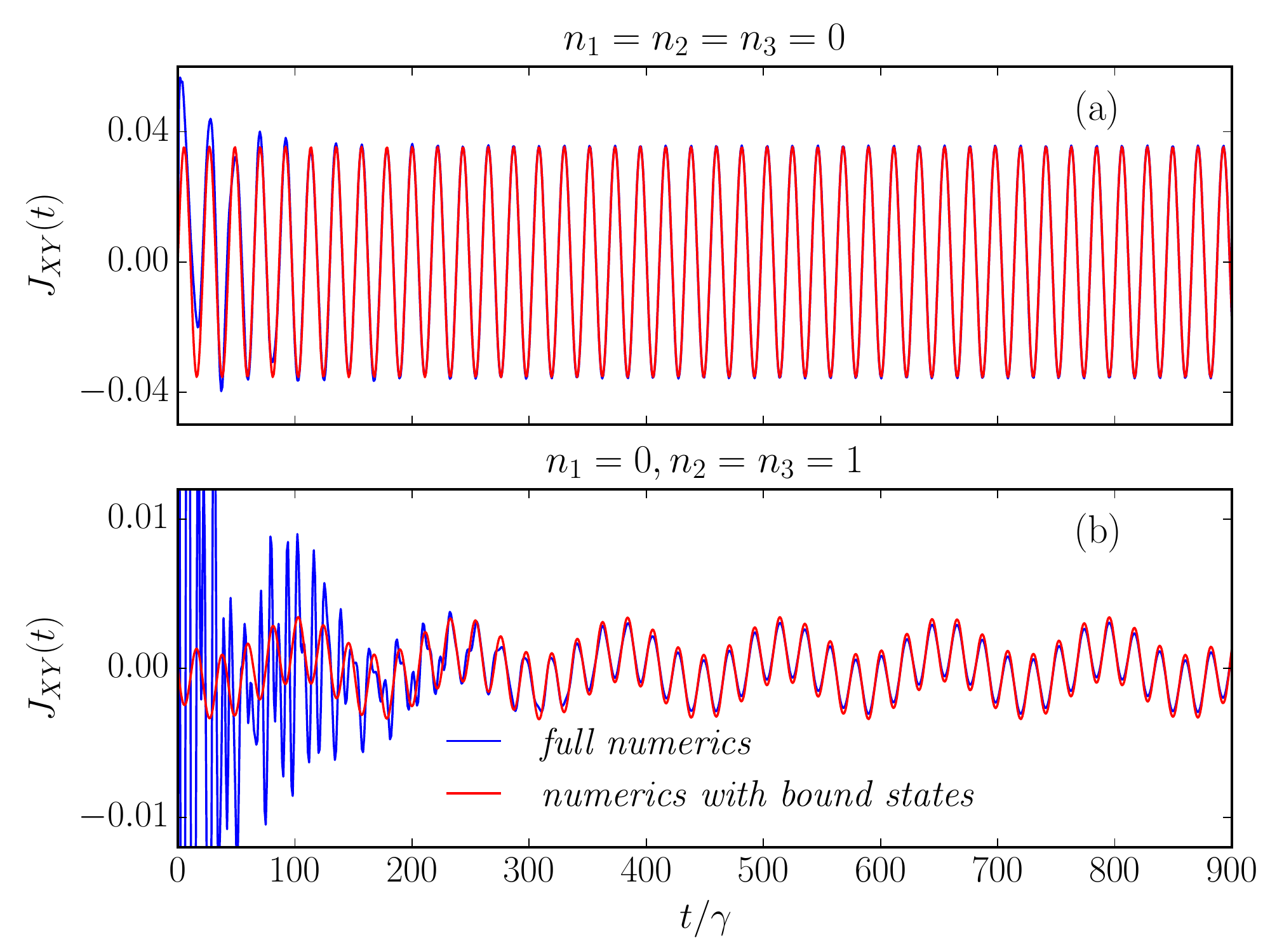}
\caption{Temporal evolution of the zero-bias current $J_{\rm XY}(t)$ in a TS-N-TS device made of identical TS wires, and the role of N wire's initial density $n_{l'}$. The blue curves are obtained from full numerics, and the red curves are from numerics with only bound states. In all panels, $L_{\rm X}=L_{\rm Z}=900,L_{\rm Y}=3$, $\gamma_{\rm X}=\gamma_{\rm Z}=1$, $\gamma_{\rm Y}=0.5$, $\Delta_{\rm X}=\Delta_{\rm Z}=0.3$, $\epsilon_{\rm X}=\epsilon_{\rm Z}=0$, $\epsilon_{\rm Y}=0.05$, $\gamma_{\rm XY}=\gamma_{\rm YZ}=0.25$, and $T_{\rm X}=T_{\rm Z}=0.02$. The above parameters (except lengths) are in units of $\gamma$.} 
\label{TopJJNw3}
\end{figure}

Next, we consider a TS-N-TS device which has been investigated for the Josephson effect in topological systems \cite{Rokhinson2012, Zazunov2012, Zazunov2016, Ioselevich2016}. We take the temperature of both the TS wires to be the same and very low, and set their chemical potential to zero. Then, one would naively expect zero electrical current in such junctions of identical TS wires at a long time. Surprisingly, we find a persistent and oscillating electrical current at both the junctions of the TS-N-TS device. In Fig.~\ref{TopJJNw3}, we show time evolution of $J_{\rm XY}(t)$ from $t_0=0$ for different initial density $n_{l'}$ of the middle N wire. The amplitude and period of the current oscillation in Fig.~\ref{TopJJNw3}(a,b) depend on the initial density of the N wire. For example, there is only a single oscillation period in Fig.~\ref{TopJJNw3}(a) for initial density, $n_1=n_2=n_3=0$, while there are two oscillation periods (a short time and a long time) in Fig.~\ref{TopJJNw3}(b) for $n_1=0,n_2=n_3=1$. Thus, our TS-N-TS device does not equilibrate. 

 The absence of unique long-time steady-state (equilibration) in a TS-N-TS device is due to the presence of bound states near the junctions. The wave functions of these bound states decay exponentially deep inside the TS wires. The energy of these bound states lies within the superconducting gap of the TS wires. Apart from the Majorana bound states of the TS wires near the intersections, there can be such bound states originating from the N wire. We observe that the amplitude and period of the current oscillation in the TS-N-TS device also depend on the total number of such bound states near the junctions \cite{Suppl}. For example, there are two Majorana bound states with energy $\pm 0.12236$ at the edges of TS wires near the intersections as well as two energy eigenstates (at energy $\pm 0.16795$) from the middle N wire inside the superconducting gap for the TS-N-TS device in Fig.~\ref{TopJJNw3}. The energy of MZMs near the junctions of the TS wires is non-zero due to an overlap of these modes through the short middle wire.  From these bound state energies, we derive the periods of short and long time oscillations respectively as $2\pi/(0.16795+0.12236)\approx 21.6$ and $2\pi/(0.16795-0.12236)\approx 137.8$ which are in good agreement with the simulated periods in Fig.~\ref{TopJJNw3}.      

\begin{figure}
\includegraphics[width=0.99\linewidth]{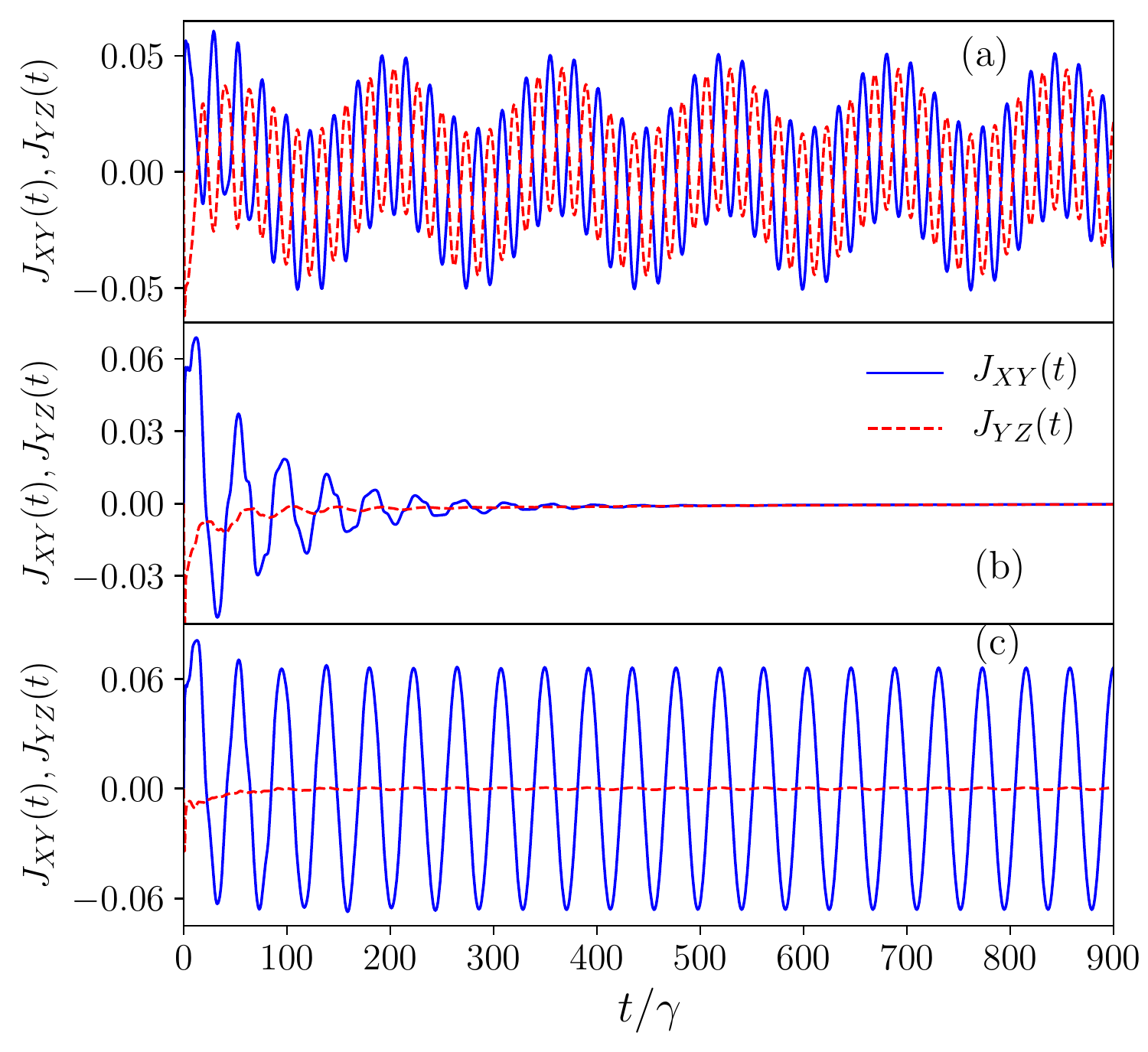}
\caption{Nature of zero-bias currents in a TS-N-Z device when the on-site energy of the superconducting Z wire is swept through a topological phase transition. The blue full lines are for $J_{\rm XY}(t)$ and the red dashed lines are for $J_{\rm YZ}(t)$. The on-site energy $\epsilon_{\rm Z}=1,2,2.5$ in panel a,b,c respectively. In all panels, $L_{\rm X}=L_{\rm Z}=900,L_{\rm Y}=3$, $\gamma_{\rm X}=\gamma_{\rm Z}=1$, $\gamma_{\rm Y}=0.5$, $\Delta_{\rm X}=\Delta_{\rm Z}=0.3$, $\epsilon_{\rm X}=0$, $\epsilon_{\rm Y}=0.05$, $\gamma_{\rm XY}=\gamma_{\rm YZ}=0.25$, $n_{l'}=0$ and $T_{\rm X}=T_{\rm Z}=0.02$. The above parameters (except lengths) are in units of $\gamma$.}
\label{TopJJNw3Vhr}
\end{figure}

To further illustrate the role of these bound states in the absence of equilibration, we study time-evolution in a TS-N-Z device where we either apply an N wire for Z or tune the on-site energy of a superconducting Z wire to sweep through a topological phase transition. We do not find any persistent, oscillating electrical current in a TS-N-N device in the absence of bias. The band of a semi-infinite N wire in our model is continuum without a bulk-gap. Therefore, the Majorana bound state of the TS wire near the junctions as well as the energy levels of the middle N wire do equilibrate with the boundary N wire; thus there is no oscillating current \cite{Suppl}.

We find persistent, oscillating currents (e.g., check Fig.~\ref{TopJJNw3Vhr}(a)) in a TS-N-Z device in the absence of voltage or thermal bias when $\epsilon_{\rm Z}<2\gamma_{\rm Z}$, such that there are Majorana bound states at the edges of the topological Z wire, and the bulk-gap ($\sim |\epsilon_{\rm Z}-2\gamma_{\rm Z}|$) in the energy spectrum of the Z wire is much larger than $\gamma_{\rm YZ}^2$ -- dissipation induced by the coupling between Y and Z wires. Again, we can separately estimate the periods of oscillating currents in  Fig.~\ref{TopJJNw3Vhr}(a) using the bound state energies, and they are $2\pi/(0.15455+0.11557)\approx 23.3$ and $2\pi/(0.15455-0.11557)\approx 161.2$. Regardless of the initial density of the middle N wire, there is no persistent, oscillating electrical current at the long time in a TS-N-Z device for $\epsilon_{\rm Z} \approx 2\gamma_{\rm Z}$ when there is either no bulk-gap in the energy spectrum or a bulk-gap which is smaller or same order of $\gamma_{\rm YZ}^2$. In Fig.~\ref{TopJJNw3Vhr}(b), we show rapid decays of $J_{\rm XY}(t)$ and $J_{\rm YZ}(t)$ to zero with time from initial time $t_0=0$ when $\epsilon_{\rm Z}=2\gamma_{\rm Z}$. We also notice in Fig.~\ref{TopJJNw3Vhr}(b) that the amplitude of $J_{\rm YZ}(t)$ is much smaller than $J_{\rm XY}(t)$ in the time duration when they are non-zero. The last fact indicates that such decays of $J_{\rm XY}(t)$ and $J_{\rm YZ}(t)$ are due to equilibration by the thermal Z wire which affects the YZ junction more than the XY junction. Finally, we show oscillating $J_{\rm XY}(t)$ and $J_{\rm YZ}(t)$ in Fig.~\ref{TopJJNw3Vhr}(c) when the Z wire is in a topologically trivial phase with a large bulk-gap for $\epsilon_{\rm Z}>2\gamma_{\rm Z}$. However, the amplitude of $J_{\rm YZ}(t)$ is almost two order smaller than $J_{\rm XY}(t)$ due to the large bulk-gap and the absence of MZM in the Z wire making the effective coupling between the Y and Z wires much smaller than that between the Y wire and the Majorana bound state in the TS wire. The pattern of oscillation in Fig.~\ref{TopJJNw3Vhr}(c) differs from that in  Fig.~\ref{TopJJNw3Vhr}(a) due to the presence of an extra Majorana bound state in the topological Z wire near the junction for Fig.~\ref{TopJJNw3Vhr}(a). The energy of MZM near the junction of the left TS wire remains almost zero due to negligible hybridization when the Z wire is topologically trivial. The period of current oscillation in Fig.~\ref{TopJJNw3Vhr}(c) is then estimated solely by the bound state energy from the middle N wire as $2\pi/0.14833 \approx 42.4$.

We learn from our above discussion for Figs.~\ref{TopJJNw3},\ref{TopJJNw3Vhr} that the persistent current oscillation in different hybrid devices in the absence of bias is due to the bound states and their initial preparation. We here show that we can reproduce the long-time persistent current oscillations in Figs.~\ref{TopJJNw3},\ref{TopJJNw3Vhr} by solely keeping the contributions of the  bound states in the current operator in Eq.~\ref{current}. To perform such calculation, we apply truncation both in real space as well as in energy eigenstates in our computation which becomes numerically inexpensive. While we keep only energy eigenstates (bound states) which are localized near the junctions, we constraint the span of real space depending on the localization lengths of the bound states around the junctions. We use these truncation rules to Eqs.~\ref{indm1},\ref{EGF}, and employ those truncated quantities along with Eqs.~\ref{indm2},\ref{indm3} in Eq.~\ref{tdm} to evaluate Eq.~\ref{current}. The calculated currents from such truncated computation with bound states are shown in Fig.~\ref{TopJJNw3} to compare them with the currents from the full simulation. They show an excellent match at long-time.     

The fractional Josephson effect with a characteristic $4\pi$ periodic current-phase relation in topological superconductors has been investigated for an unambiguous detection of MZMs \cite{Kitaev2001,LutchynPRL2010,Kwon2004,Rokhinson2012,ChuanNature2018}. There are also many tunneling spectroscopy studies with N and superconducting tip to probe various magnetic and non-magnetic bound states in conventional and unconventional superconductors \cite{NadjPergeScience2014,RubyPRL2015,Yazdani1999,Hudson2001,JiPRL2008}. Our present dynamical study can in principle be tested in both these above set-ups. Our finding of lack of equilibration in the presence of bound states in a device with superconducting boundary wires with a bulk-gap poses challenge for direct detection of dc or ac fractional Josephson effect as well as electrical current measurements in such device. We conclude that the necessary condition for equilibration (or a unique steady-state) in these hybrid junctions is either the absence of bound states near the junctions or the band of one or both the boundary wires being continuum without a bulk-gap.    

\begin*{\it Acknowledgments.}
We are grateful to A. Dhar and D. Sen for discussion. NB acknowledges funding from DST-FIST programme. DR acknowledges funding from the DST, India via the Ramanujan Fellowship. 
\end*

\bibliography{bibliography1}

\clearpage

\newpage
\vspace{3cm}
\onecolumngrid
\begin{center}
{\large \bf Supplemental Material for Dynamics of hybrid junctions of Majorana wires} \\
\vspace{2mm}
{Nilanjan Bondyopadhaya and Dibyendu Roy}
\end{center}
\vspace{3mm}
\twocolumngrid
\setcounter{figure}{0}
\renewcommand\thefigure{S\arabic{figure}}
\setcounter{equation}{0}
\renewcommand\theequation{S\arabic{equation}}
\label{Suppl}  
\section{N-TS-N device}
\begin{figure}
\includegraphics[width=\columnwidth]{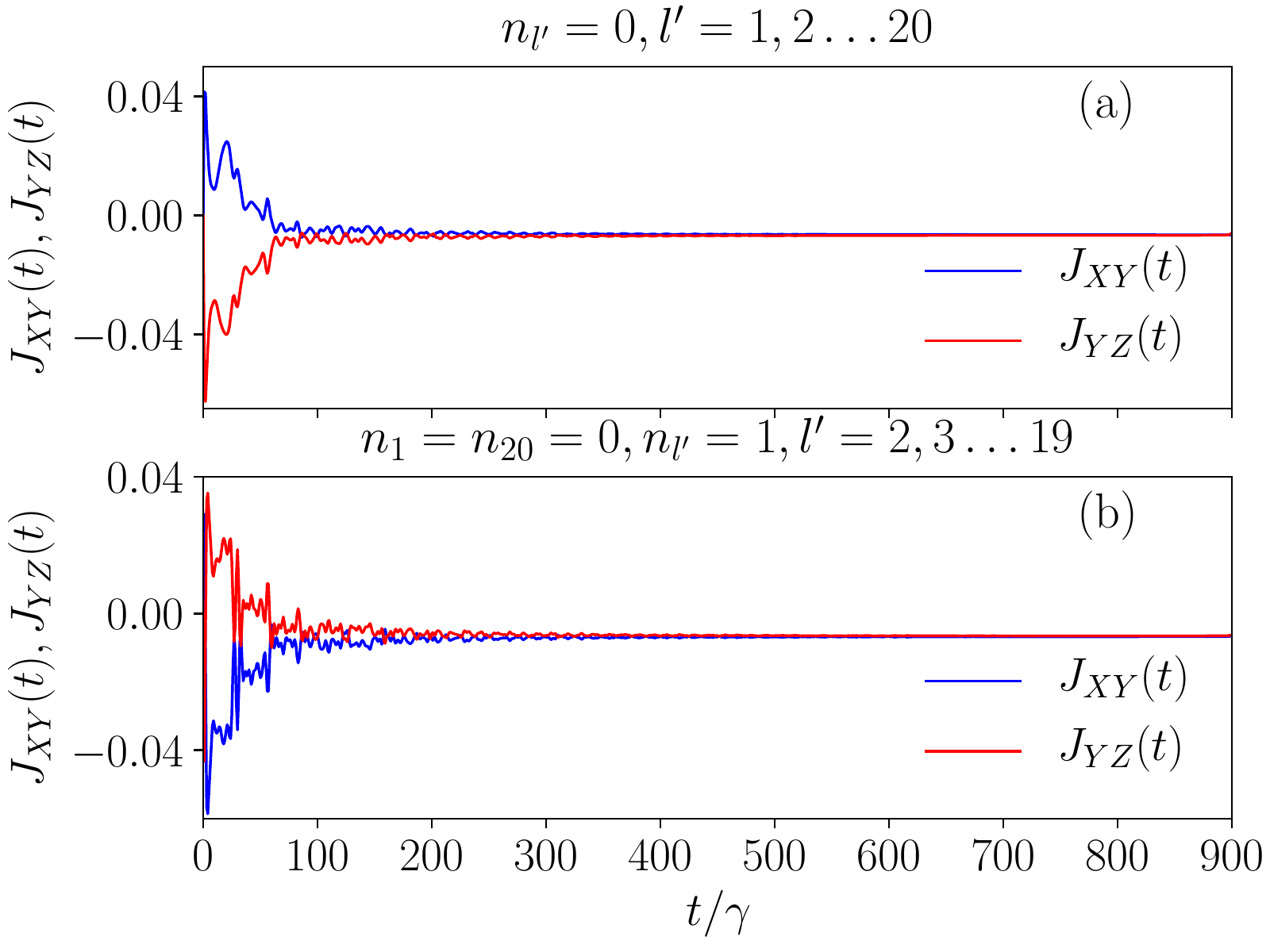}
\caption{Temporal evolution of the currents in an N-TS-N device under a voltage bias and presence of a unique nonequilibrium steady-state. The initial density of the TS wire is indicated on the heading of the panels. In all panels, $L_{\rm X}=L_{\rm Z}=900,L_{\rm Y}=20$, $\gamma_{\rm X}=\gamma_{\rm Z}=1$, $\gamma_{\rm Y}=0.5$, $\Delta_{\rm X}=\Delta_{\rm Z}=0$, $\Delta_{\rm Y}=0.1$, $\epsilon_{\rm X}=\epsilon_{\rm Y}=\epsilon_{\rm Z}=0$, $\gamma_{\rm XY}=\gamma_{\rm YZ}=0.25$, $T_{\rm X}=T_{\rm Z}=0.2$ and $\mu_{\rm X}=-0.2,\mu_{\rm Z}=0.2$. All above parameters except lengths are in units of $\gamma$.}
\label{NTSN}
\end{figure}

In the main text, we have mentioned that the nonequilibrium steady-state transport in an N-TS-N device seems to be independent of the initial conditions of the finite TS wire when the band of the N wires is wider than that of the TS wire. In Fig.~\ref{NTSN}, we present plots to support our claim. In Ref.~\cite{Vasseur2014}, the time-evolution of the many-electron wave function of an N-TS junction is studied after a quantum quench in which the N and TS wires of equal and extended lengths are connected at some initial time. The decoupled N and TS wires are initially prepared in their respective ground states, and the overlap between the time-evolved wave function with the initial state (the Loschmidt echo) is found to decay with a universal power law in time for large times after the quench. We here instead prepare the decoupled N wires of our N-TS-N device initially in thermal equilibrium. In Fig.~\ref{NTSN}, we show that the electrical currents at a long time  do not depend on the initial density of the middle TS wire. We find from our numerics that the currents at both the junctions reach a constant non-zero value when there is a voltage bias across the TS wire due to the chemical potential difference in the boundary N wires. The simulated steady-state current in Fig.~\ref{NTSN} from time-evolution of the Heisenberg equations of motion matches with that from the Fourier transform solution at steady-state \cite{RoyPRB2012, RoyPRB2013}.

\section{Longer length of N wire in TS-N-TS device}
\begin{figure}
\includegraphics[width=0.99\linewidth]{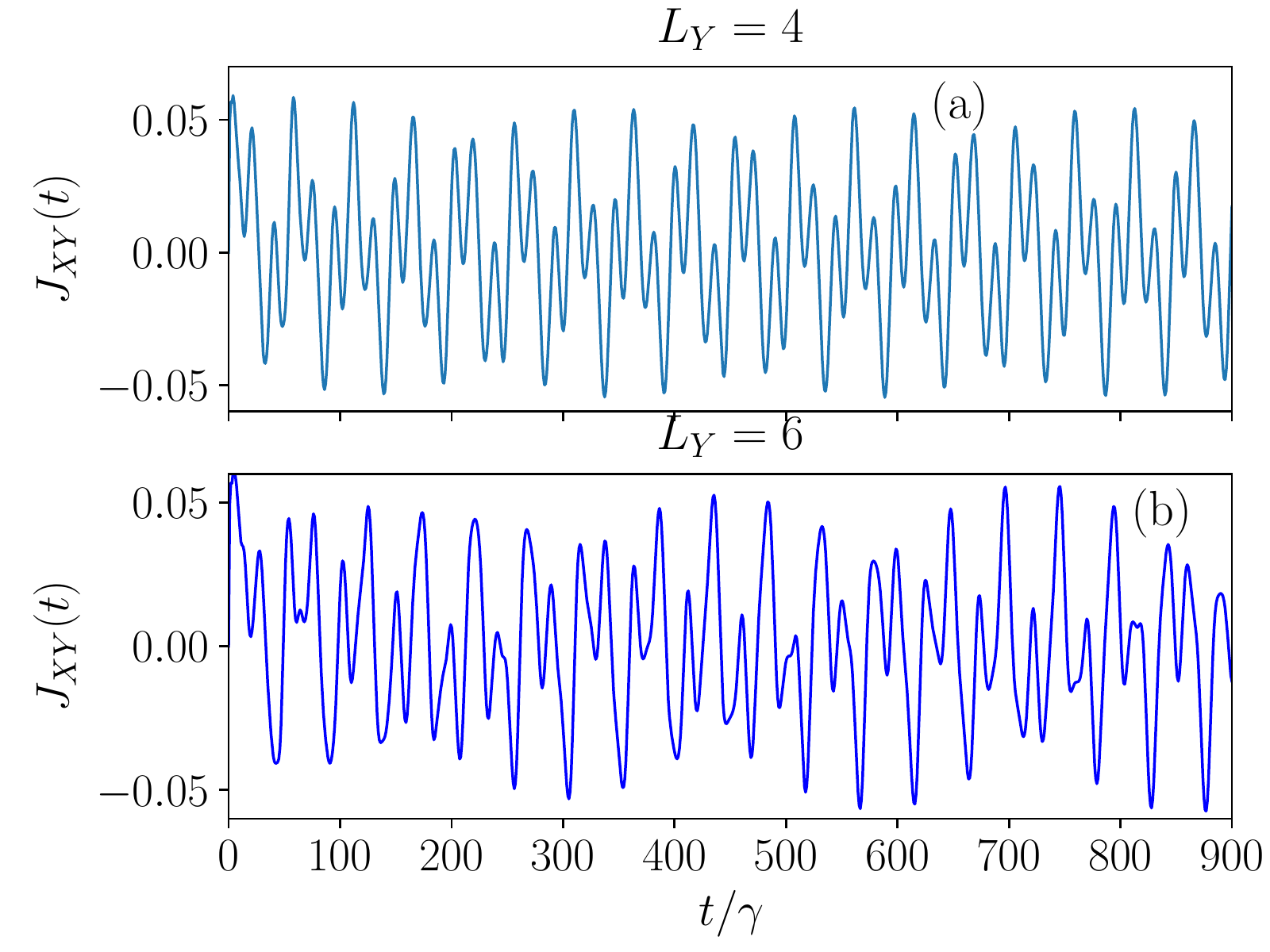}
\caption{Temporal evolution of the zero-bias current $J_{\rm XY}(t)$ in a TS-N-TS for different length of the middle N wire. The length of the middle N wire is indicated on the heading of the panels. In all panels, $L_{\rm X}=L_{\rm Z}=900,L_{\rm Y}=4,6$, $\gamma_{\rm X}=\gamma_{\rm Z}=1$, $\gamma_{\rm Y}=0.5$, $\Delta_{\rm X}=\Delta_{\rm Z}=0.3$, $\epsilon_{\rm X}=\epsilon_{\rm Z}=0$, $\epsilon_{\rm Y}=0.05$, $\gamma_{\rm XY}=\gamma_{\rm YZ}=0.25$, $n_{l'}=0$ and $T_{\rm X}=T_{\rm Z}=0.02$. All above parameters except lengths are in units of $\gamma$.} 
\label{TSnTS}
\end{figure}

In Fig.~\ref{TSnTS}, we show how the amplitude and period of the current oscillation in a TS-N-TS device change with an increasing length of the middle N wire. The number of bound states from the middle N wire can increase with increasing wire length. The current oscillations in the TS-N-TS device seem to survive for relatively long N wire length. We have seen finite persistent current oscillation in the junctions for $L_{\rm Y}=50$. However, the pattern of current oscillation, e.g., oscillation period, becomes a bit complicated in the presence of multiple bound states for a longer $L_{\rm Y}$. We present in Fig.~\ref{TSnTS} plots for $L_{\rm Y}=4,6$ for zero initial density of the middle N wire. 

\section{TS-N-N device}
\begin{figure}
\includegraphics[width=\columnwidth]{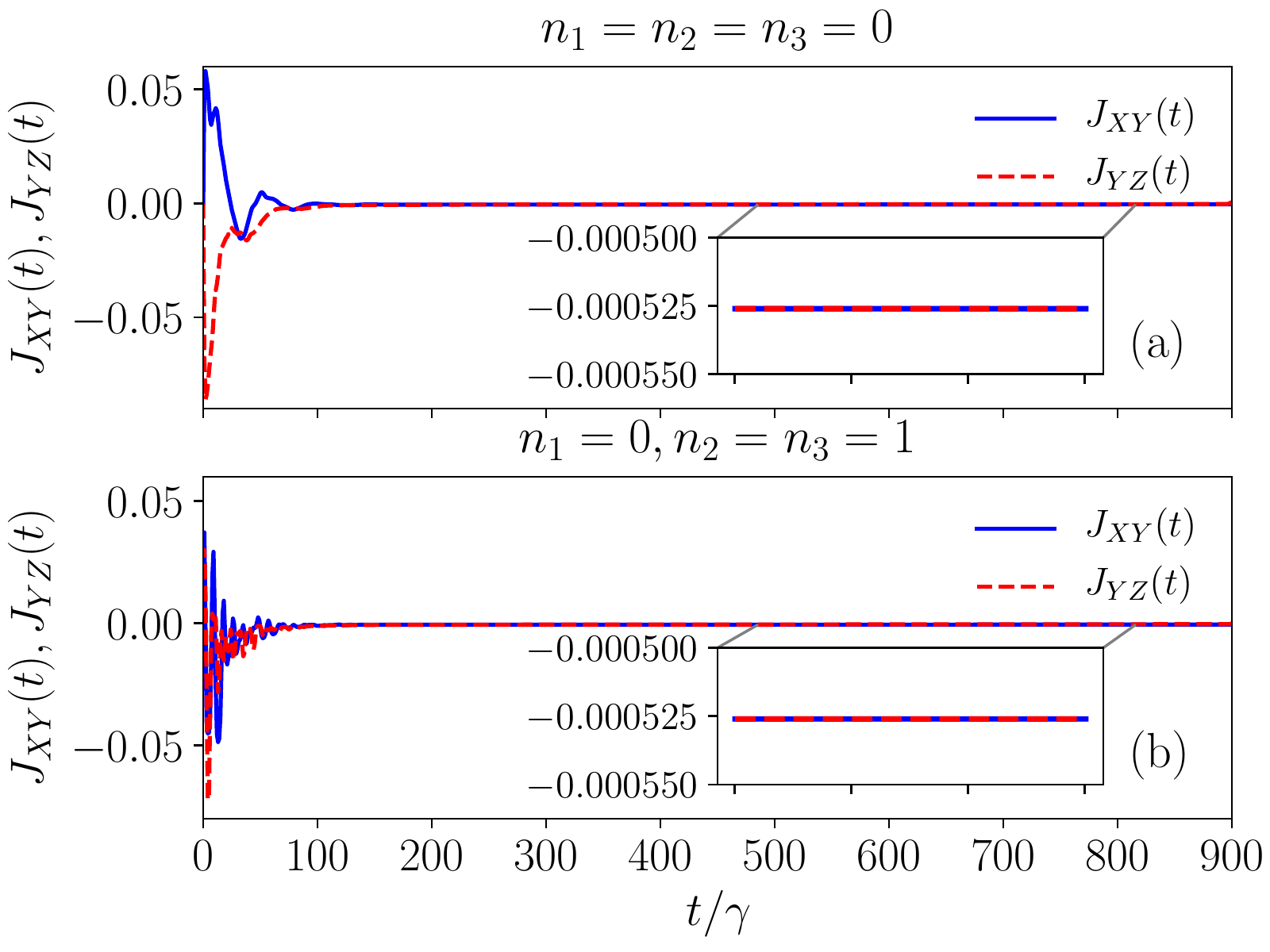}
\caption{Temporal evolution of the currents in a TS-N-N device under a voltage bias and presence of a unique nonequilibrium steady-state. The initial density of the middle N wire is indicated on the heading of the panels. In all panels, $L_{\rm X}=L_{\rm Z}=900, L_{\rm Y}=3$, $\gamma_{\rm X}=\gamma_{\rm Z}=1$, $\gamma_{\rm Y}=0.5$, $\Delta_{\rm X}=0.2,\Delta_{\rm Y}=\Delta_{\rm Z}=0$, $\epsilon_{\rm X}=0.02, \epsilon_{\rm Y}=0.05, \epsilon_{\rm Z}=0$, $\gamma_{\rm XY}=\gamma_{\rm YZ}=0.25$, $T_{\rm X}=T_{\rm Z}=0.2$ and $\mu_{\rm X}=0,\mu_{\rm Z}=0.5$. All above parameters except lengths are in units of $\gamma$.}
\label{TSnN}
\end{figure}

We have got unique nonequilibrium steady-state in a TS-N-N device, and we do not see here any persistent electrical current oscillation in the absence of voltage or temperature bias. In Fig.~\ref{TSnN}, we plot electrical currents at both the junctions of a TS-N-N device for two different initial density in the middle N wire. We also apply a voltage bias across the middle N wire by using a non-zero chemical potential for the boundary N wire. We find that the electrical currents at both junctions are independent of the initial density in the middle N wire. The nonequilibrium steady-state currents are also the same at both intersections. Therefore, we conclude that the middle N wire in a TS-N-N device equilibrates with the boundary wires due to a continuous band of the boundary N wire.

\section{Role of tunnel coupling}
\begin{figure}
\includegraphics[width=0.99\linewidth]{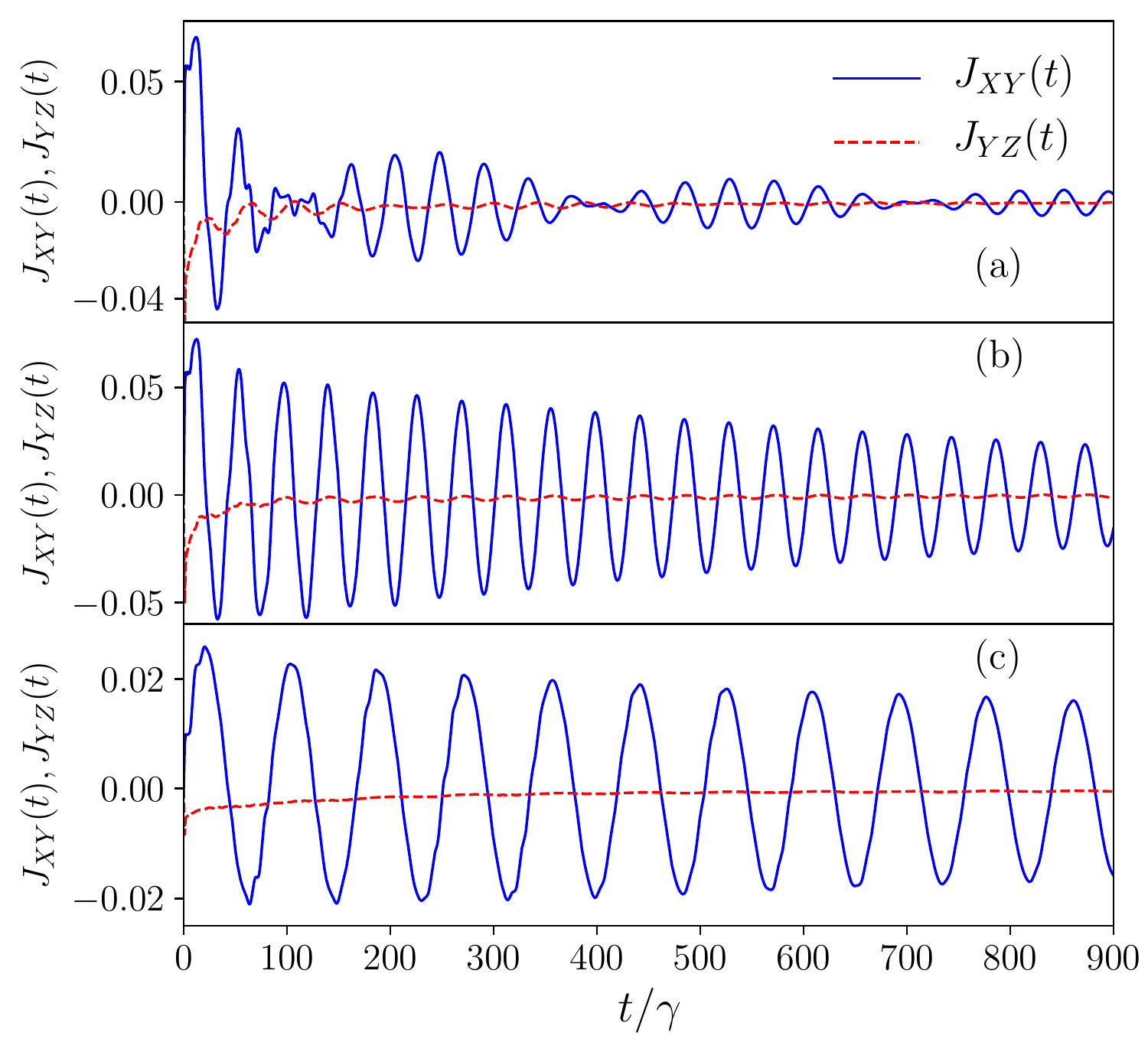}
\caption{Temporal evolution of the currents in a TS-N-Z device when the on-site energy of the superconducting Z wire is swept through the topological phase transition, and role of tunnel coupling. The blue full lines are for $J_{\rm XY}(t)$ and the red dashed lines are for $J_{\rm YZ}(t)$. The on-site energy $\epsilon_{\rm Z}=1.99,2.05,2.05$ and tunnel coupling $\gamma_{\rm YZ}=0.25,0.25,0.1$ in panel a,b,c respectively. In all panels, $L_{\rm X}=L_{\rm Z}=900,L_{\rm Y}=3$, $\gamma_{\rm X}=\gamma_{\rm Z}=1$, $\gamma_{\rm Y}=0.5$, $\Delta_{\rm X}=\Delta_{\rm Z}=0.3$, $\epsilon_{\rm X}=0$, $\epsilon_{\rm Y}=0.05$, $\gamma_{\rm XY}=0.25$, $n_{l'}=0$ and $T_{\rm X}=T_{\rm Z}=0.02$. The above parameters (except lengths) are in units of $\gamma$.}
\label{Tunnel}
\end{figure}

In the main text, we have discussed equilibration in a TS-N-Z device when the Z wire transits through a topological phase transition. The closing of superconducting bulk-gap near the topological phase transition is the reason for equilibration. In Fig.~\ref{Tunnel}, we further show that the electrical current in a TS-N-Z device decays to zero at a long time when there is a bulk-gap in the Z wire which is the smaller or same order of $\gamma_{\rm YZ}^2$. The tunnel coupling $\gamma_{\rm YZ}$ induces broadening of the energy of the bound states. Thus, the energy-broadened bound states near the junctions can overlap with the band of the Z wire when the bulk-gap is small. Therefore, we expect equilibration of bound states and decay of electrical currents when the ratio between the bulk-gap and $\gamma_{\rm YZ}^2$ is low. For example, this ratio is 0.16, 0.8, 5 respectively for three plots in Fig.~\ref{Tunnel}(a,b,c). While the current $J_{\rm XY}(t)$ in Fig.~\ref{Tunnel}(a,b) show clear decay with time, the decay of $J_{\rm XY}(t)$ is relatively slow in Fig.~\ref{Tunnel}(c). Therefore, a stronger tunnel coupling can effectively reduce the bulk-gap in superconducting boundary wires, and hence can help in equilibration. Nevertheless, there can also be other mechanisms of equilibration such as electron-phonon interaction at the middle wire or irradiation of radiation  on the Josephson junctions for the measurement of Shapiro steps in current-voltage curves. The bound states can equilibrate when the phonon or the radiation at the junctions is wide-band such that the band of phonon or radiation closes the superconducting gaps.  

\section{Truncated numerics with bound states}
We have seen in the main text that the bound states play a crucial role in the persistent electrical current oscillation in different hybrid junctions in the absence of bias. These bound states may be formed inside the middle wire as well as at the edges of the boundary TS wires. To study the contribution of these bound states to the time evolution of density matrix, let us first rewrite the full Green's function $\mathcal{G}^+_{r,s}(t-t_0)$ in the following way \cite{DharPRB2006}:
\bea
&&\mathcal{G}^+_{r,s}(t-t_0)= \mathcal{G}^{b+}_{r,s}(t-t_0)+\mathcal{G}^{c+}_{r,s}(t-t_0) \nn \\
&=&-i\sum_{b} \Psi_{b}(r) \Psi_{b}^*(s) e^{- i \Lambda_{b} (t-t_0)}
-i \int d\xi \rho^c_{rs}(\xi)e^{-i \xi (t-t_0)} \, ,\nn \\
\label{EGF1}
\eea
where $\Psi_b$ and $\Lambda_b$ are the full system's eigenvectors and eigenvalues corresponding to the bound states localized around the junctions, and the density matrix $\rho_{rs}^c$ is given by a sum over extended (continuum) states $ \Psi_{q}$ of the system $\rho^c_{rs} (\xi)=\sum_{q}^{c} \Psi_{q}(r) \Psi_{q}^*(s)\delta(\xi-\Lambda_q)$. Clearly, $ \mathcal{G}^{b+}(t-t_0)$ includes all the contribution from the bound states of the full system, while $\mathcal{G}^{c+}(t-t_0)$ incorporates the contribution of the continuum states. If the system does not support bound states, the first term becomes zero and the second term controls the steady-state behavior of the system. However, the scenario changes drastically in the presence of bound states in the system. 

Since the bound states are localized, the magnitude of the elements of $\mathcal{G}^{b+}(t-t_0)$ becomes small if we move away from the spatial regions where the bound states are localized. This property of $\mathcal{G}^{b+}(t-t_0)$ is very useful for our computational purpose. Suppose, we know that all the bound states are localized within the region $\mathcal{R}_0$. % demarcated by site indices $l'=X$ to $l'=X+N$.
Then it is legitimate to choose the elements of $\mathcal{G}^b(t-t_0)$ to be zero outside the region $\mathcal{R}_0$, and we calculate the elements $\mathcal{G}^{b+}(t-t_0)$ in the region $\mathcal{R}_0$ only. Although the boundary of $\mathcal{R}_0$ is not unique, one can extract the essential features of $\mathcal{G}^{b+}(t-t_0) $ even with a choice of region $\mathcal{R}_0$ which merely encompasses the spatial extent of all the overlapping bound states.

In our numerics for studying persistent current oscillation, we consider a TS-N-TS device where both the boundary TS wires have the same chemical potential and temperature. Moreover, we also kept both the superconducting wires in the topologically non-trivial phase away from the topological phase transition. For such a system, there exist two Majorana bound states at the ends of each TS wires. If the TS wires are taken to be sufficiently long, the overlap of the Majorana modes from opposite ends of an individual TS wire becomes almost negligible. However, the overlap between the Majorana modes from the right end of left boundary TS wire and the left end of right boundary TS wire remains finite as long as the length of the middle N wire is not very large. Further, there may also exist bound states within the middle N wire, and it can have non-zero overlap with the Majorana modes localized at the inner edges of the boundary TS wires. In such a situation, essential behavior of the currents in the junctions can be studied with the bound state dependent part of the Green's function, i.e., $\mathcal{G}^{b+}(t-t_0)$. As it has been argued already, this $\mathcal{G}^{b+}(t-t_0)$ is chosen to be zero for the region outside $\mathcal{R}_0$ where $\mathcal{R}_0$ is extended from $L_X-\delta$ to $L_X+L_Y+\delta$ with $L_Y$ is the length of the N wire, $L_X$ is the length of left boundary TS wire and $\delta$ ($ << L_X,\, L_Y,\, L_Z$) is the number of sites inside the boundary wires. The value of $\delta$ solely depends on the spatial extent of bound state wave-functions inside the boundary wires. For example, if the wave-functions of the bound states around the junctions have non-zero amplitude up to the $L_\delta$ number of sites inside the boundary wires from the junction edge, then $\delta \approx L_\delta$. For our simulation, we find that it is sufficient to choose $\delta \backsim 10$ for the boundary wires with size $L_{X(Z)} \backsim 10^3$ and the middle wire length $L_Y \backsim 10$.

Given the above discussion, it may be possible to study the time evolution of the bound state dependent part of the density matrix using the following expression:
%\lipsum[1]
\begin{widetext}
\bea
\langle a^\dagger_l(t) a_m(t) \rangle_b &=& \sum_{r, s = 2(L_X-\delta) }^{2 L_X} \mathcal{G}^{b+}_{m,s }(t-t_0) \langle a^\dagger_{r}(t_0) a_{s}(t_0) \rangle [\mathcal{G}^{b+}_{l,r}(t-t_0)]^\dagger+ \sum_{r,s=2L_X+1}^{2(L_X+L_Y)} \mathcal{G}^{b+}_{m,s}(t-t_0) \langle a^\dagger_r(t_0) a_s(t_0) \rangle [\mathcal{G}^{b+}_{l,r}(t-t_0)]^\dagger \nn \\ &&+\sum_{r, s=2(L_X+L_Y)+1}^{2(L_X+L_Y+ \delta )} \mathcal{G}^{b+}_{m,s}(t-t_0) \langle a^\dagger_{r}(t_0) a_{s}(t_0) \rangle [\mathcal{G}^{b+}_{l,r}(t-t_0)]^\dagger,
\label{denmatB}
\eea
\end{widetext}
where $\langle a^\dagger_{r}(t_0) a_{s}(t_0) \rangle$ in the first and third term denotes the initial equilibrium density matrix of the left and right boundary wire respectively, and $\langle a^\dagger_{r}(t_0) a_{s}(t_0) \rangle$ in the second term represents the initial density matrix of the middle wire. This simplified analytic expression is quite useful in studying the characteristic features of the full density matrix, $\eta_{lm}(t)=\langle a^\dagger_l(t) a_m(t) \rangle$, provided the system hosts bound states with non-zero overlap among themselves. However, if the system does not possess bound states, or the bound states have negligible overlap, the expression (\ref{denmatB}) may fail to depict the actual dynamics of $\eta(t)$ as the continuum part of the full Green's function ($ \mathcal{G}^{c+}(t-t_0) )$ starts to play a dominant role. If  Eq.~\ref{denmatB} fails, then we need to restore to the original expression in Eq.~\ref{tdm} of the main text for studying the time evolution; even though it is numerically expensive.

Presence of Majorana bound states at the edges of the boundary TS wires helps us to further simplify the analytical expression for $\langle a^\dagger_l(t) a_m(t) \rangle_b$ in some cases. To this end, let us write the initial equilibrium density matrix of the boundary wires as a summation of two contributions,
\begin{widetext}
\bea
\langle a^{\dagger}_{r}(t_0) a_{s}(t_0) \rangle &=& \langle a^{\dagger}_{r}(t_0) a_{s}(t_0) \rangle_{b'}+\langle a^{\dagger}_{r}(t_0) a_{s}(t_0) \rangle_c \nn \\
&& =\sum_{b'} {\psi_{b'}^\alpha}^* (r) \psi_{b'}^\alpha (s) f(\lambda_{b'}^\alpha,\mu_\alpha, T_\alpha)+\sum_{c} {\psi_{c}^\alpha }^* (r) \psi_{c}^\alpha(s) f(\lambda_c^\alpha,\mu_\alpha, T_\alpha),
\label{dmb}
\eea
\end{widetext}
where $\psi_q^\alpha $ and $\lambda^\alpha_q$ are the eigenvectors and eigenvalues of the boundary wires $\alpha=X,\, Z$ (see Eq.~\ref{bathham} of the main text), and $r,s=2(L_{\rm X}-\delta),\dots, 2L_{\rm X}$ for $\alpha={\rm X}$, and $r,s=2(L_{\rm X}+L_{\rm Y})+1, \dots, 2(L_{\rm X}+L_{\rm Y}+\delta)$ for $\alpha={\rm Z}$. In the second line of Eq.~\ref{dmb}, the first term represents the contribution of Majorana bound states localized at the edges of boundary TS wires in topologically non-trivial phase, and the second term denotes the participation of the continuum states of the boundary wires. Here $b'$ and $c$ subscripts designate the bound states and the continuum states of the boundary TS wires respectively. 

Substituting \ref{dmb} in \ref{denmatB}, we get 
\begin{widetext}
\bea
%\eta_{m,l}(t)=
&&\langle a^\dagger_l(t) a_m(t) \rangle_b = \sum_{r, s = 2(L_X-\delta) }^{2 L_X} \mathcal{G}^{b+}_{m,s }(t-t_0) \langle a^\dagger_{r}(t_0) a_{s}(t_0) \rangle_{b'}\, [\mathcal{G}^{b+}_{l,r}(t-t_0)]^\dagger+ \sum_{r,s=2L_X+1}^{2(L_X+L_Y)} \mathcal{G}^{b+}_{m,s}(t-t_0) \langle a^\dagger_r(t_0) a_s(t_0) \rangle [\mathcal{G}^{b+}_{l,r}(t-t_0)]^\dagger \nn \\ &&+\sum_{r,s=2(L_X+L_Y)+1}^{2(L_X+L_Y+ \delta )} \mathcal{G}^{b+}_{m,s}(t-t_0) \langle a^\dagger_{r}(t_0) a_{s}(t_0) \rangle_{b'}\, [\mathcal{G}^{b+}_{l,r}(t-t_0)]^\dagger+\sum_{r,s = 2(L_X-\delta) }^{2 L_X} \mathcal{G}^{b+}_{m,s }(t-t_0) \langle a^\dagger_{r}(t_0) a_{s}(t_0) \rangle_{c}\, [\mathcal{G}^{b+}_{l,r}(t-t_0)]^\dagger \nn \\
&&+\sum_{r,s=2(L_X+L_Y)+1}^{2(L_X+L_Y+ \delta )} \mathcal{G}^{b+}_{m,s}(t-t_0) \langle a^\dagger_{r}(t_0) a_{s}(t_0) \rangle_{c}\, [\mathcal{G}^{b+}_{l,r}(t-t_0)]^\dagger,
\label{denmatF}
\eea
\end{widetext}
where the initial density matrix in the first and third term depends on the bound states of the boundary wires, that of the second term relies entirely on the initial conditions at the middle wire, whereas the initial density matrix in the fourth and final term depends on the continuum states of the boundary wires. 
For the sake of brevity, we can further rewrite the Eq. \ref{denmatF} in the following form 
\beq
\langle a^\dagger_l(t) a_m(t) \rangle_b=\langle a^\dagger_l(t) a_m(t) \rangle_b^B+\langle a^\dagger_l(t) a_m(t) \rangle_b^c, 
\label{denmatRB}
\eeq
where $\langle a^\dagger_l(t) a_m(t) \rangle_b^B$ represents first three terms at the right hand side of Eq. \ref{denmatF}, and $\langle a^\dagger_l(t) a_m(t) \rangle_b^c$ represents rest of the terms at the right hand side of the same equation.

Interestingly, we observe that the temporal evolution of the zero-bias current in a TS-N-TS device is controlled by the initial density of the middle N wire, the density of the Majorana bounds states formed at the inner edges of the boundary wires and the bound states of the full system. Therefore, the junction currents in Eq.~\ref{current} can be approximated as 
\bea
J_{\rm XY}(t)& \approx &-2\gamma_{\rm XY}{\rm Im}[\langle a^{\dg}_{2L_X+1}(t)a_{2L_X-1}(t) \rangle_b^B],\label{currBXY}\\
J_{\rm YZ}(t)& \approx &-2\gamma_{\rm YZ}{\rm Im}[\langle a^{\dg}_{2(L_X+L_Y)+1}(t)a_{2(L_X+L_Y)-1}(t) \rangle_b^B].\label{currBYZ}\nn\\
\eea
It can be noted that the above approximation remains valid as long as the boundary wires are in topologically non-trivial phase away from the phase boundary. However, near the topological phase transition point, continuum states of TS wires contribute substantially to the junction currents; thus the above approximation fails.

Strong initial density dependence of junction currents in a TS-N-TS device can be explained with the help of Eq. \ref{EGF1}, \ref{denmatF}. For a uniform initial density $n_{l'}$ of the middle N wire, the amplitudes of the oscillating modes solely depend on the wave functions of the bounds states of the full system. Thus, the initial density of the N wire can be treated as a constant factor. However, the amplitudes of the oscillating modes are also affected by the initial densities for a  non-uniform initial density at the middle wire. As the wave functions of different bound states are also spatially non-uniform, the combined effect of wave functions of bound states and inhomogeneous initial density results in suppression of the amplitudes of some selected frequency modes and (or) enhancement of others in the persistent current oscillations.
So, the overall time-evolution dynamics of the density matrix behaves very differently than that for a uniform initial density.

\end{document}